\documentclass[11pt]{article} 

\usepackage[utf8]{inputenc} 

\usepackage{geometry} 
\usepackage{graphicx} 

\usepackage{booktabs} 
\usepackage{array} 
\usepackage{paralist} 
\usepackage{verbatim} 
\usepackage{subfig} 

\usepackage{fancyhdr} 
\usepackage{sectsty}
\allsectionsfont{\sffamily\mdseries\upshape} 

\usepackage[nottoc,notlof,notlot]{tocbibind} 
\usepackage[titles,subfigure]{tocloft} 

\usepackage{color}
\usepackage{epsfig}
\usepackage{euscript}
\usepackage{mathrsfs}
\usepackage{amsmath}
 \usepackage{relsize}
\allowdisplaybreaks[1]

\title{\bf A Route toward Quantum Gravity through the Imaginary-Time Field Theory}
\author{Yi-Cheng Huang\\
E-mail: {\tt ychuang1109@msn.com}
}

\begin{document}
\maketitle
{\bf Abstract}
\vskip.2cm
A possible way toward the quantization of a  weak gravitational field inspired by the imaginary-time field theory  is discussed.  The analogies of the general relativity in the canonical formulation with the thermodynamic geometry and superfluids are also pointed out. The proposed regularization method in the imaginary-time formalism of field theory shows  a good chance removing theoretical obstacles and constructing a perturbative theory of quantum gravity.    
\newpage

\section{Introduction}
One of the main issues in physics over several decades is the quantization of the gravitational field.  The Standard Model has successfully quantized three of  the four fundamental interactions and gives predictions with a high precision except the gravitation. The obstruction is coming from that the quantization of the Einstein-Hilbert action is not renormalizable \cite{renormalizablity}. In general, we may decide whether a quantum field theory is renormalizable or not by its superficial degree of divergence, $\mathcal{D}$. When $\mathcal{D}>0$, the theory is non-renormalizable, and $\mathcal{D}=4$ from the  Einstein-Hilbert action at one-loop level. In order to absorb  divergent terms from the multi-loop radiative corrections, an infinite number of bare Lagrangian other than the curvature scalar $R$ have to be added into the action, for example, $R^2$ and $R^{\mu\nu} R_{\mu\nu}$ for one-loop level. As a result, many of the efforts on the quantum gravity have turned to non-perturbative and background-independent methods. In the perturbative approach, the gravitational field is separated into two parts: $g^{\mu\nu}(x)=\eta^{\mu\nu}_{\rm b.g.}(x)+h^{\mu\nu}(x)$. The background metric $\eta^{\mu\nu}_{\rm b.g.}(x)$ may not only be flat, it could be a curved space-time. One of the background-independent approaches is the loop quantum gravity (LQG) \cite{loopqg}. A smooth background field is thought of as the cause of the UV divergence, and a theory of quantum gravity without the separation is called background-independent.  In LQG, space is quantized into a granular structure.  The connections among the loops are called spin network, and the worldsheet of a spin network is called spin foam, whose time evolution is controlled by Hamiltonian constraints. One of the difficulties for LQG to become a candidate theory of the quantum gravity was to give the general relativity in the correspondence limit, and the newly proposed master constraint \cite{thiemann}, classically equivalent to an infinite number of the Hamiltonian constraints, could yield a hope establishing the semi classical limit. In addition, LQG is successful in deriving the entropy of non-singular black holes in accordance with Hawking's formula. Recently, the measurements of the quantum gravity are begun to be considered by physicists in astrophysical, cosmological  observations and gravitational wave detectors. \par 
There are  other attempts regarding the theory of quantum gravity, such as the effective field theory of the quantum gravity \cite{effectiveqg}, canonical quantum gravity \cite{canonicalqg}, entropic gravity \cite{jacobson}, string theory \cite{string}, etc.\,. In the following, the canonical quantum gravity will be shown to have many analogies in theory with the thermodynamical geometry and superfluids in Section \ref{hamiltonian}. In Section \ref{advantage}, the thermodynamical aspects of the gravity will be emphasized and its connection with the imaginary-time field theory will be discussed. The new regularization method, which is proposed in \cite{huang13a}, will be explored for a possible theory of quantum gravity.

\section{Implications from Hamiltonian Approach }
\label{hamiltonian}

\subsection{Hamiltonian of Einstein-Hilbert action}
The conceptual difficulty between the general relativity and  the quantum theory is the role that time plays. In general relativity, time is dynamic, while in quantum mechanics it is absolute. As time is elevated to be a dynamic variable, it is found there is no role that time can play in the theory. This is called the problem of time \cite{time} in quantum gravity, and it will be shown in the canonical formulation. The Einstein-Hilbert action in the general relativity  reads
\begin{eqnarray}
S_{\rm E.H.}=\int\frac{d^4 x}{16\pi G}\sqrt{-g}\,R,\label{einsteineq}
\end{eqnarray}
where $R$ is the  scalar curvature, $g$ is the determinant of the metric tensor, $g_{\mu\nu}$, and $G$ is the gravitational constant with a dimension of [mass]$^{-2}$. In the Hamiltonian formalism \cite{canonicalqg} it can be expressed in the form of
\begin{eqnarray}
S_{\rm E.H.}=\frac{1}{16\pi G}\int dt d^3x\left(p^{ab}\dot{h}_{ab}-N\mathcal{H}^{\rm g}_\perp-N^a\mathcal{H}^{\rm g}_a\right),\label{einsteineqinh}
\end{eqnarray}
where the canonical momenta $p^{ab}\equiv \frac{\partial \mathcal{L}^{\rm g}}{\partial \dot{h}_{ab}}$; with the alphabetic index goes from 1 to 3, 
\begin{eqnarray*}
\mathcal{H}^{\rm g}_\perp={16\pi G} G_{abcd} p^{ab}p^{cd}-\frac{\sqrt{h}}{16\pi G}\left(^{(3)}R-2\Lambda\right),\,\,{\rm and}\,\,\,\mathcal{H}^{\rm g}_a=-2D_b p^{\,b}_a,
\end{eqnarray*}
where $h$ is denoted as the determinant of the 3-metric $h_{ab}$, $^{(3)}R$ is the extrinsic curvature scalar, $\Lambda$ is the cosmological constant, $N$ is the lapse function, $N^a$ is the shift vector and the so-called (inverse) DeWitt metric is defined as $G_{abcd}\equiv \frac{1}{2\sqrt{h}}\left(h_{ac}h_{bd}+h_{ad}h_{bc}-h_{ab}h_{cd}\right)$.  The operator $D_b$  is the covariant derivative with respective to the coordinate $x^b$. The line element is in the form 
\begin{eqnarray}
ds^2=(h_{ab}N^aN^b-N^2)dt^2+2h_{ab}N^a dx^b dt+h_{ab}dx^a dx^b,\label{adm}
\end{eqnarray}
which is called the Arnowitt-Deser-Misner (ADM) decomposition of the metric. The Hamiltonian can be identified from eq. (\ref{einsteineqinh}), and it reads 
\begin{eqnarray}
H^{\rm g}\equiv \int d^3 x \mathcal{H}^{\rm g}\equiv\int d^3x\left(N\mathcal{H}^{\rm g}_\perp+N^a\mathcal{H}^{\rm g}_a\right).\label{einsteinham}
\end{eqnarray}
In eq. (\ref{einsteineqinh}), the lapse function $N$ and the shift vector $N^{a}$ can be regarded as the Lagrange multipliers for the constraints $\mathcal{H}^{\rm g}_\perp\doteq0$ and $\mathcal{H}^{\rm g}_a\doteq0$, the sign $"\doteq"$ indicates the equals sign required by the constraint. This thus results in the problem of time. As the traditional time $t$ changes its role to a dynamical variable and becomes a function $t(\tau)$ of an absolute time variable $\tau$, a constraint, like $\mathcal{H}^{\rm g}_a\doteq0$, in quantization of the action in the general relativity gives rise to a timeless condition, $\hat{H}(x)|\psi\rangle=0$, the so-called Wheeler-DeWitt equation \cite{wheelerdewitt}, thus time does not substantially take part in the theory.

\subsection{Ruppeiner, Weinhold metric and thermodynamical length}
\label{ruppeiner}
A unique point of view on the thermodynamics developed from the angle of Riemann geometry starts to create some possibilities in understanding gravity. Early works on the study of the thermodynamics by using the geometrical methods were devoted by Gibbs \cite{gibbs} and Caratheodory \cite{caratheodory}. The thermodynamic metrics proposed by Weinhold \cite{weinhold} and Ruppeiner \cite{ruppeiner} are defined as
\begin{eqnarray}
g^{\rm W}_{ij}=\partial_i\partial_j U (S,X^a),&{\rm and}&g^{\rm R}_{ij}=-\partial_i\partial_j S (U,X^a),
\end{eqnarray}
where $U$ is the internal energy, $S$ is the entropy, and $X^a$ refers to the extensive parameters of the system.
The two may be related by $ds^2_{\rm R}={ds^2_{\rm W}}/{T}$, where $T$ is the temperature of the system. On the other hand, the partition function can be found to define the metric through the knowledge of  the so-called thermodynamical length \cite{weinhold,ruppeiner,thermodynamiclength}, which is to measure the distance between equilibrium thermodynamic states. Consider a thermodynamical system, which is in equilibrium with a large thermal reservoir, the probability distribution is rendered through
\begin{eqnarray}
p\left(x|\lambda\right)=\frac{1}{Z}e^{-\beta H(x,\lambda)}= \frac{1}{Z}e^{-\sum_i\lambda^i(t) X_i(x)},
\end{eqnarray}
where $H(x,\lambda)$ is the Hamiltonian and is split into several collective variables, $X_i(x)$. An example is shown in eq. (\ref{einsteinham}) for ${H}^{\rm g}$ in the general relativity. The variables $X_i(x)$ are $\mathcal{H}^{\rm g}_\perp$ and $\mathcal{H}^{\rm g}_a$, and $\lambda_i$ are the Lagrange multipliers, the lapse function $N$ and shift vector $N_a$. The partition function $Z$ is related to the Gibbs free energy $F$, the Massieu potential $\psi$, and entropy $S$:
\begin{eqnarray}
\ln Z=-\beta F=\psi=S-\sum_i\lambda^i\langle X_i\rangle,
\end{eqnarray}
where $\langle ...\rangle$ indicates the average over an equilibrium ensemble. The thermodynamic metric tensor can be given from the second derivative of the Massieu potential 
\begin{eqnarray}
g_{ij}=\frac{\partial^2\psi}{\partial\lambda^i\partial\lambda^j}=\langle (X_i-\langle X_i\rangle)(X_j-\langle X_j\rangle)\rangle.\label{thermometric}
\end{eqnarray}
If the index $i$ is associated with spatial or time coordinate, the  matrix may be related to the spacetime metric. This could be the case of $N$ and $N_a$ for the Hamiltonian of the Einstein-Hilbert action. On the other hand, this can make a connection with the theory in information geometry, which  applies the techniques of differential geometry to the field of probability theory. The Fisher  information matrix defined in the information geometry gives
\begin{eqnarray*}
g_{ij}(\lambda)&=&\sum_x p(x)\frac{\partial \ln p(x)}{\partial \lambda^i}\frac{\partial \ln p(x)}{\partial \lambda^j},\\
&=&\sum_x p(x)\left(X_i+\frac{\partial \psi}{\partial \lambda^i}\right)\left(X_j+\frac{\partial \psi}{\partial \lambda^j}\right),\\
&=&\langle (X_i-\langle X_i\rangle)(X_j-\langle X_j\rangle)\rangle.
\end{eqnarray*}
This is the same with the result of eq. (\ref{thermometric}). Here only the analogies and the correspondences in the theoretical structures among the general relativity and other fields are presented in a brief way. More details can be found in the cited references.

\subsection{Superfluid analogies}
\label{superfluid}
\begin{figure}  
\begin{center}  
\includegraphics[height=10cm,width=14cm,angle=0]{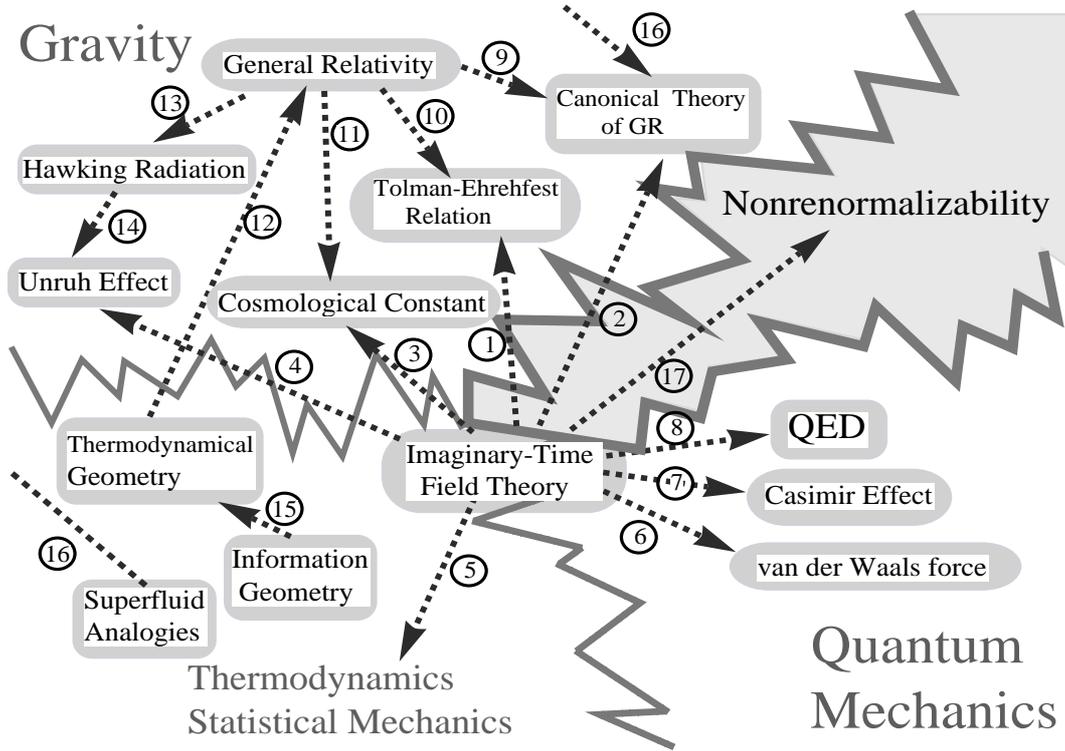}  
\caption{\small  The gray crack indicates the nonrenormalizability of GR, which is the reason why the gravity can not be quantized. (1) The two can be connected through the scale transformation. (2) Canonical theory of GR in the imaginary-time field theory may shed some light on the quantum gravity. 
(3) The imaginary-time field theory predicts a finite value of the cosmological constant \cite{huang13c}. (4) The imaginary-time field theory is able to make an agreeable result of the Unruh effect. (5) The imaginary-time field theory is intrinsically a field theory of finite temperature for vacuum. (6), (7) The imaginary-time field theory predicts agreeable results of the Casimir effect and van der Waals force, and automatically carries regularization functions for them to remove divergences. \cite{huang13b} (8) The imaginary-time field theory make agreeable one-loop radiative correction results  and renormalization group equations with QED. (9) The Canonical theory of GR is the Hamiltonian formulation of the general relativity. (10) A perfect fluid or radiation in a gravitational field described by the general relativity obeys the Tolman-Ehrenfest relation, $T\sqrt{g_{00}}=const.$\,.  (11) In the cosmological observations, the term of the cosmological constant added in the Einstein equation is needed to account for the acceleration of the universe. (12) Thermodynamic geometry uses the Riemann geometry, which is the language of the general relativity, to study the thermodynamics. (13) Hawking radiation is a black-body radiation emitted from the event horizon of the black hole. (14) Unruh temperature has the same form as the Hawking temperature. (15) Information geometry is a mathematics using the differential geometry to study the probability theory, and thermodynamic geometry is one of its branches. (16) Superfluid has an effective Riemann metric similar to the ADM decomposition of the metric in the canonical theory of GR. (17) The imaginary-time field theory has a new regularization method inspired from the scale transformation and is expected to solve the difficulty of the nonrenormalizability in the perturbative quantum gravity.  
 \label{fig:ARQGITTgraph}}  
\end{center}  
\end{figure}  

Due to the ultraviolet divergences at small scales, it may indicates that the quantum field theory is not a fundamental theory and both of it and the general relativity are effective theories, which are not applicable at all scales. Thus if one may find analogies of the quantum field theory from some well known fields, like condensed matter theory, it may help us to understand the physics at trans-Planckian scale, since in principle we know the condensed matter structure at any relevant scale. An example is the superfluid $^3$He-A \cite{volovik01} ($^3$He in phase A), which in the limit $T\rightarrow 0$ it gradually acquires all of the symmetries, including Lorentz invariance, local gauge invariance, etc., from nothing.
The Lagrangian of the spin-wave, which is described by $\alpha$, the angle of the antiferromagnetic vector, in antiferromagnets and in $^3$He-A takes the form 
\begin{eqnarray}
\mathcal{L}&=&\mathsmaller{\frac{1}{2}}\eta^{ij}\nabla_i\alpha\nabla_j\alpha-\mathsmaller{\frac{1}{2}}\chi\left(\dot{\alpha}+({\bf v}\cdot\nabla)\alpha\right)^2=\mathsmaller{\frac{1}{2}}\sqrt{-g}g^{\mu\nu}\partial_\mu\alpha\partial_\nu\alpha,
\end{eqnarray}
where the matrix $\eta^{ij}$ is the spin rigidity, $\chi$ is the spin susceptibility and ${\bf v}$ is the local velocity of crystal in antiferromagnets. The superfluid velocity is also equal to ${\bf v}$. The components of the effective Riemann metric can be written as 
\begin{eqnarray}
g_{00}=-\mathsmaller{\frac{1}{\sqrt{\eta\chi}}}\left(1-\chi\eta_{ij}v^iv^j\right), \,g_{oi}=-\sqrt{\frac{\chi}{\eta}}\eta_{ij}v^j, \,g_{ii}=\sqrt{\frac{\chi}{\eta}}\eta_{ij},\,\,{\rm and}\,\,\sqrt{-g}=\sqrt{\frac{\chi}{\eta}},
\end{eqnarray}
where $\eta$ is the determinant of $\eta_{ij}$. This form of metric corresponds to the ADM decomposition of the metric in eq. (\ref{adm}) in the canonical theory of the Einstein-Hilbert action, and the corresponding lapse function is $N=(\eta \chi)^\mathsmaller{-\frac{1}{4}}$. Of course, it would be interesting to see if there  exist analogous functions of $\eta$ and $\chi$ for the general relativity. There are also analogies of black holes in quantum liquids \cite{volovik03}. The imaginary-time field theory and its relationships with other fields of physics  and physical effects are illustrated in Figure \ref{fig:ARQGITTgraph}.

\section{Advantages from the Imaginary-Time Formalism}
\label{advantage}
\subsection{Why imaginary time?}
In quantum field theory, the partition function, $Z[J]$, a generating functional, is used to describe the physics involved with fields, such as electrons, photons, etc.\,. With this treatment of quantization, instead of the way in the quantum mechanics, we are also able to derive the propagators of the scalar and spinor fields, and the interactions between various fields are treated perturbatively in this approach. On the other hand, in the statistical mechanics, the partition function, $Z$, which sums over all of the possible states, describes the statistical properties of a system in thermodynamic equilibrium. Its main difference from the first  partition function  is that the countable collection of the  variables, such as discrete quantum numbers, is replaced by the an uncountable set of the functional of a field $\phi$.  However, the two partition functions show a great similarity, by changing the time variable in $Z[J]$ to the imaginary-time of a finite range, the two expressions may transform from one to another. In the condensed matter physics, Green functions of nonzero temperature are derived from the partition function with the  imaginary-time, $\tau$, ranging from 0 to $\beta$, and are expanded by the Matsubara frequency. One important  property induced by the imaginary-time formalism is the KMS condition \cite{kms}. The Green functions of finite temperature are found to be cyclic with respective to the variable $\tau$, $G_{\rm F}(\tau)=-G_{\rm F}(\tau+\beta)$  for fermions, and $G_{\rm B}(\tau)=G_{\rm B}(\tau+\beta)$  for bosons. In addition, the general relativity is perceived to have thermodynamical properties, such as the Tolman-Ehrenfest relation \cite{tolman30}, which states the relation between the metric and the temperature of a fluid in the gravitational field. Another one from these properties to which physicists have paid much attention is the Hawking radiation.  As an insufficient but interesting argument, the Hawking temperature can be derived  
from the Schwarzschild metric with the variable transform, $t=-i\tau$. The imaginary-time $\tau$ behaves like a polar angle in the polar coordinate system, and the period of the polar angle thus yields the Hawking temperature, $T=\frac{1}{8\pi GM}$ \cite{zee}. It can be expected if the thermodynamical meanings are considered in the Einstein-Hilbert action in the imaginary-time and space, this euclidean Schwarzschild solution might mean more than it appears.


\subsection{Agreements and differences from the conventional theories}
\subsubsection{Agreements and implications}
Many physical effects may attribute to vacuum, including the Lamb shift \cite{lamb}, the Casimir effect \cite{casimir},  van der Waals force \cite{vanderwaals}, the Unruh effect \cite{unruh}, the Hawking radiation \cite{hawking}, the dark energy \cite{darkenergy}, etc.\,. From the theoretical side, in QED the Lamb shift can be explained by the radiative corrections between a proton and an electron in the hydrogen atom. In the language of the QED, there is no obvious thermal properties for vacuum and it takes the renormalization to remove the UV divergence in order to extract results with physical meanings. From the classical electromagnetism,  the Casimir effect can be predicted by summing the discrete energies of all of the available electromagnetic waves between two conducting plates. While with a half classical and half quantum approach, the main part of the van der Waals force is described by the interaction between instantaneous multi-poles in molecules. Classical electromagnetism apparently has no  thermodynamical descriptions for the vacuum, and the calculations of the Casimir effect and the van der Waals force also face  crises of divergence. Intentionally added regularization functions help to produce finite results.  An astonishing prediction for a vacuum is the Unruh effect, which states that an accelerating observer will detect  black-body radiation from empty space. The temperature of the thermal bath is called the Unruh temperature. A similar counterpart to the Unruh effect is the Hawking radiation, which predicts a black hole radiates near its event horizon also with a black-body spectrum. Though the derivation of the Hawking radiation does not take a theory of quantum gravity, only the knowledge of the field theory in curved space-time is needed, it is recognized that it gives some clues about what properties for a theory of quantum gravity could have.  In the modern cosmology, the dark energy is identified as another face of the vacuum and is described by a cosmological constant in the Einstein equation. Theoretically, the  cosmological constant is computed by the zero-point energy of the vacuum; however it brings the largest discrepancy of  the vacuum energy \cite{weinberg89} as compared with the experimental value. Another approach to the cosmological constant in the field theory is the DeWitt-Schwinger representation \cite{schwinger51,dewitt75}; unfortunately, it also leads to a divergent quantity and needs a bare cosmological constant to make the final result finite. \par
In the work of \cite{huang13a}, a theory of QED in the the imaginary-time formalism is developed. Many interesting features are worth more emphases and exploration. The first is that the UV divergences in a Feynman integral at temperature $T\,(=\frac{1}{\beta})$ are canceled by another integral at an extremely high temperature $T_0\, (=\frac{1}{\beta_0})$, as the integration domain of the imaginary-time is an closed-open interval, $(0^+,\beta\,]$. The integral of $\beta_0\rightarrow 0^+$ is yielded from the lower bound of the $\tau$-integration domain. Thus, an extra integral accompanies every Feynman integral and takes out the divergent part. This overturns the criteria of a viable field theory that needs to be renormalizable, or say the necessity of enough bare terms in the Lagrangian, since in the imaginary-time field theory no counter term are necessary. Thus we don't have to worry about running out of the counter terms anymore. Secondly, the radiative corrections to one-loop level in the imaginary-time QED are proved to be in agreement with the conventional QED, such as the Lamb shift, etc.  One thing to notice is that, even though it is the imaginary-time field theory, there are two types of the propagators; one is for the imaginary-time and the other is for the real-time. The Feynman integral from the imaginary-time propagator leads to the real part of a conventional Feynman integral, while the real-time propagator gives rise to the imaginary part. Thus it shows there are mathematical correspondences between the new and the conventional field theories. The third is the renormalization group effect. In the dimensional regularization, the renormalization scale $\mu$ is inevitably introduced and  describes the  concomitant variations in the couplings, masses and field strengths. On one hand, the actions in the imaginary-time formalism are required to have scale invariance; on the other hand, the scale transformation is often used to express the electromagnetic waves  in the field theory of the curved space-time, and the Tolman-Ehrenfest relation \cite{tolman30}  indicates the metric is related to the temperature of the fluid filled in space. It agrees with the formalism in the field theory of the imaginary-time for the Matsubara frequency is temperature-dependent. All of the above clues points to that the temperature determines the physical scale, just like the renormalization scale $\mu$. Thus there is no wonder that, in the imaginary-time field theory, the same renormalization group equations as in $\overline{MS}$ renormalization scheme  are recovered with respect to the change of the temperature. Besides, the exploration of the vacuum effects mentioned in the first paragraph through the imaginary-time formalism  not only demonstrates great compatibilities with the precedent theoretical predictions but also suggests many implications \cite{huang13b}. The thermalized theory of vacuum for the Casimir effect and the van der Waals force automatically provides regularization functions, which are added intentionally in the old approaches, to remove the redundant divergences through the formalism of the Matsubara frequency. On the other hand, the thermalized vacuum renders an entity of heat bath for the Unruh effect. The imaginary-time Green functions not only satisfies the descriptions of the thermal field in the Unruh effect and thus gives the same prediction of the black-body spectrum, but also agrees with those of the vacuum in the Hawking radiation. In the recent work \cite{huang13c}, a finite cosmological constant is able to be predicted, based on the imaginary-time formalism, in two approaches along with the implementation of the scale transformation. One is the DeWitt-Schwinger representation, which yields a divergent result in the conventional field theory, and the other is  following the calculation of the Casimir effect and taking limit of the distance between the two plates to the size of the universe. Each gives a correct ratio of the equation of state, $w=-1$, which is an important criteria for the dark energy. Besides, the largest energy discrepancy  between the experimental and the theoretical values of vacuum energy can be conciliated in the theoretical structure of the imaginary-time field theory. Hence, it would be interesting to see if this structure is able to give more clues for the quantum gravity.\par
In addition, though the vacuum effects are so feeble for detections, there have been some experimental data to prove their existence. The Casimir effect has been confirmed through different kinds of experimental configurations \cite{mohideen,bressi}, and recently its dynamic version for the moving mirror \cite{paraoanu13} has also been verified. And of course, the van der Waals force is a well known effect in thermodynamics. The so-called Sokolov-Ternov effect \cite{steffect} has been found as a special case of the Unruh effect.


\subsubsection{A  new regularization method and a hope of quantum gravity}
In the work of \cite{huang13a}, a new regularization method from a perspective of thermodynamics is proposed. 
As briefly described in the previous section, the regulation  and the cancellation of the UV divergences are made possible through taking  advantages of the finite integration domain of the imaginary-time and the scale invariance of the actions. An analogy of this technique is the  Casimir effect, where the Casimir force can be derived from the difference between the continuous and the discrete potentials \cite{casimir}. The mode number in the direction normal to the two conducting plates is only allowed by the viable standing waves in between and its momentum $\frac{2\pi n}{L}$, where $L$ is the distance of the two plates, becomes discrete. This is similar to the proposed technique, at a very high temperature, that is equal to a very small $\beta_0$, the discrete Matsubara frequency $\frac{2\pi n}{\beta_0}$ stands out of the rest continuous dynamical variables. As a consequence, finite results with physical meanings are coming from taking the difference of the continuous and discrete radiative corrections. 
The advantage given by the imaginary-time formalism is that no counter term in the Lagrangian is needed. As we recall the difficulty of quantizing the Einstein-Hilbert action is that it is notoriously non-renormalizable. 
For instance, the UV divergences  from one-loop radiation corrections of gravitons can not be absorbed completely into  the counter terms provided by the bare Einstein-Hilbert Lagrangian, and the situation gets worse for higher order corrections.
 Conceptually, according to the viewpoints  that have been discussed: the gravity seems to have a thermodynamical nature, thus a quantum field theory in the imaginary-time formalism may look like a possible solution. As a result, the imaginary-time field theory is hoped to provide a chance for another step in pursuing a successful theory of quantum gravity, at least perturbatively. Let us take a  look back at the perturbative quantum gravity and see how the non-renormalizability can be improved.  The metric, the gravitational field, involved in (\ref{einsteineq}) can be expanded as $g_{\mu\nu}(x)=\eta_{\mu\nu}(x)+h_{\mu\nu}(x)$, where $\eta_{\mu\nu}$ is the background field and is set to the matrix $\eta_{\mu\nu}={\rm diag}(1,-1,-1,-1)$ for a flat space-time background. The second term, $h_{\mu\nu}(x)$, is the  weak gravitational field and will be quantized in a similar manner to those fields in the quantum field theory.  Expand the action in powers of $h_{\mu\nu}$, including the scalar curvature, $R$, and the Jacobian $\sqrt{-g}$, we may obtain that the action is in the form of 
\begin{eqnarray*}
S_{\rm E.H.}=\int\frac{d^4 x}{16\pi G}\left(\partial h\partial h+h\partial h\partial h+h^2\partial h\partial h+...\right),
\end{eqnarray*}
where $h$ and $\partial$ are the abbreviations of $h_{\mu\nu}$ and $\partial_\mu$. After rescaling the gravitational field, $h^{\mu\nu}\rightarrow\sqrt{G}\,h^{\mu\nu}$, it becomes
\begin{eqnarray*}
S_{\rm E.H.}=\int{d^4 x}\left(\partial h\partial h+\sqrt{G}h\partial h\partial h+Gh^2\partial h\partial h+...\right),
\end{eqnarray*}
where the rescaled $h^{\mu\nu}$ now has a dimension of [mass] instead of none. The first term in the parenthesis is the kinetic term, and the rest are for interactions.
 One of the one-loop examples of being non-renormalizable for the perturbative quantum gravity is a 
Feynman diagram, like $\raisebox{-3mm}{\psfig{figure=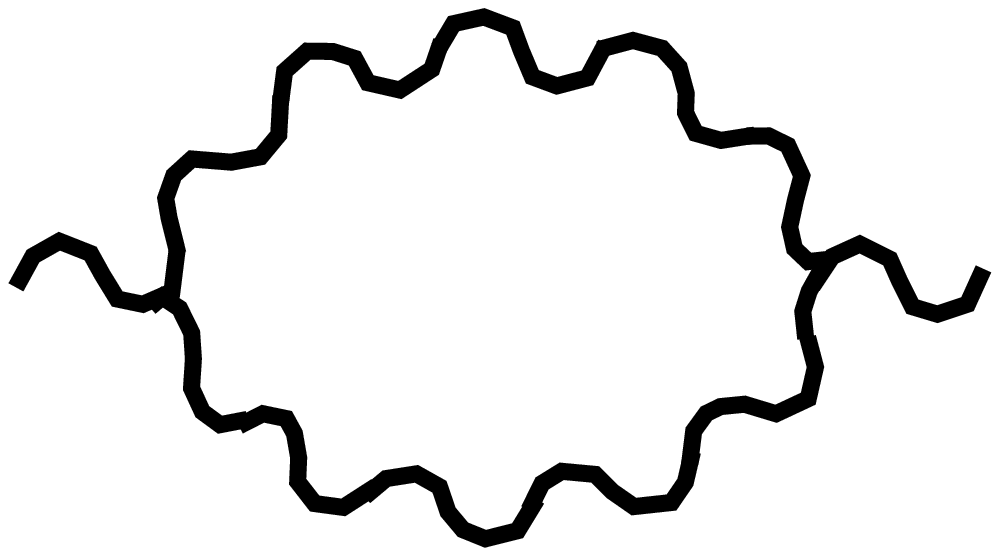, width=1.5cm,height=0.8cm}}$\,. The conventional Feynman integral takes the form $\int d^4k \frac{kkkk}{k^2k^2}$, and it is a quartic divergence. To write it down in a general form under the imaginary-time formalism, after some operations like the Feynman parametrization, it reads
\begin{eqnarray*}
\frac{1}{\beta}\sum_n\int \frac{d^3{\bf k}}{(2\pi)^3}\frac{k_n^4}{(k^2_n+\Delta)^2}=\frac{8\pi}{3}\frac{N^4_{\rm max}}{\beta^4}-\frac{2}{\pi}\frac{N^2_{\rm max}}{\beta^2}\Delta+\frac{3}{8\pi^2}\Delta^2\ln N_{\rm max}+{\rm finite\,\,terms},
\end{eqnarray*}
where $\Delta$ is a function of external momenta and masses. The cutoff of the 3-momentum is $\Lambda_{\rm cutoff}=\frac{2\pi N_{\rm max}}{\beta}$, where $N_{\rm max}$ is the maximal mode number of the Matsubara frequency. On the other hand,  as we may remember from the imaginary-time formalism, for every Feynman integral there goes another one with the limit of $\beta\rightarrow 0$, where we will denote it as $\beta_0$.  The accompanied loop integral at the limit of $\beta_0\rightarrow 0$ is
\begin{eqnarray*}
\frac{1}{\beta_0}\sum_n\int \frac{d^3{\bf k}_0}{(2\pi)^3}\frac{k_{0,n}^4}{(k^2_{0,n}+\Delta_0)^2}=\frac{8\pi}{3}\frac{N^4_{\rm max}}{\beta_0^4}-\frac{2}{\pi}\frac{N^2_{\rm max}}{\beta_0^2}\Delta_0+\frac{3}{8\pi^2}\Delta_0^2\ln N_{\rm max},
\end{eqnarray*}
where $k_{0,n}$ is the imaginary-time loop momentum for the scale at $\beta_0$.  The function $\Delta_0$ has the scale transformation with its counterpart at $\beta$: $\Delta_0=\left(\frac{\beta}{\beta_0}\right)^2\Delta$. After taking into account the scaling factor $\left(\frac{\beta_0}{\beta}\right)^4$ due to  the scaling factors given from those in the outside of the loop integral (two from the external fields and two from the coupling constants), the divergent parts can be canceled. Practical calculations regarding QED with this regularization method are performed in ref. \cite{huang13a}, as well as some examples from $\phi^3$- and $\phi^4$-theory. One thing that needs to be noticed is that the above two integrals are obtained by integration over the 3-momentum first then the integration or the summation over the Matsubara frequency; otherwise they are not comparable on the same basis. Another one-loop example, such as $\raisebox{-4mm}{\psfig{figure=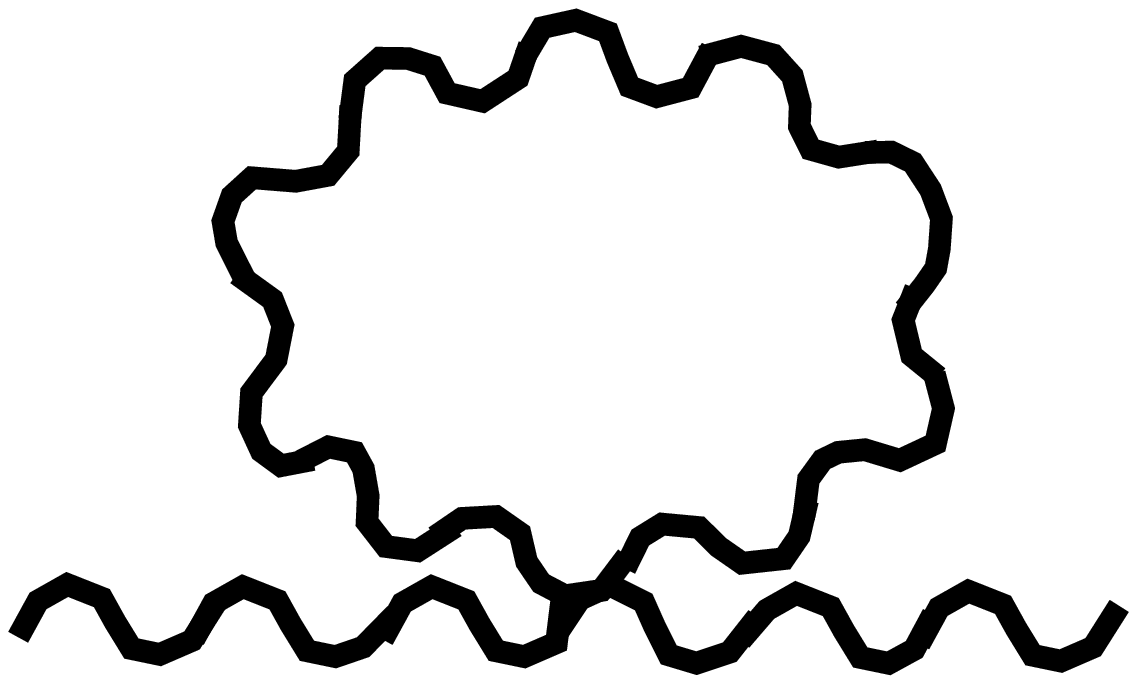, width=1.5cm,height=1cm}}$, has a conventional Feynman integral $\int d^4k \frac{kk}{k^2}$, it can be easily proved that the exact cancellation can be fulfilled by the two imaginary-time Feynman integrals. 

\section{Discussion}
In this article, the quantization of gravity using the imaginary-time field theory is discussed. Two possible approaches are presented, one is based from the canonical theory of the general relativity and the other is the perturbation theory. As for the first, the canonical theory, or say the Hamiltonian formalism, of the general relativity has shown many similarities in the theories of thermodynamic geometry and information geometry, and many analogies can also be found in some condensed matter systems, such as superfluids. On the other hand, for the perturbation theory, from the aspects in the effective field of quantum gravity, the infinite set of parameters due the non-renormalizability of the Einstein-Hilbert action is neglected in the low-energy regime, and the theory is expected no longer to be predictable in the Planck energy scale.  In the imaginary-time formalism of the field theory, the temperature is to determine the energy scale, thus different effective field theories corresponds to different degrees of vacuum temperature. According to the newly proposed regularization method, it provides an once-and-for-all way to eliminate the redundant UV divergences. Till now, it has shown agreeable results with QED and various vacuum effects and demonstrated a conciliated theoretical structure with the field theories in the curved space-time and the thermodynamical perspectives in general relativity. Thus this approach could suggest a hopeful route  in searching for a theory of quantum gravity.


\end{document}